\documentclass[reprint,aps,twocolumn,superscriptaddress,eqsecnum,pre]{revtex4} 
 
\usepackage{scrextend}
\usepackage{amssymb}  
\usepackage{amsmath}   
\usepackage{natbib}
\usepackage{epsfig}   
\usepackage{relsize}
\usepackage{graphicx,subfigure}   
\usepackage{dcolumn}
\usepackage{bm}  
\usepackage[english]{babel}
\usepackage[latin1]{inputenc}
\usepackage{hyperref}
\usepackage{xcolor}
\usepackage{comment}
\usepackage[update]{epstopdf}
\begin{document}

\title{Single-species fragmentation: the role of density-dependent feedbacks}

\author{V. Dornelas} \email{vivian@aluno.puc-rio.br}
\affiliation{Department of Physics, PUC-Rio, Rua Marqu\^es de S\~ao Vicente, 225, 22451-900, Rio de Janeiro, Brazil}
\author{E. H.  Colombo}
\affiliation{IFISC (CSIC-UIB), Campus Universitat Illes Balears, 07122, Palma de Mallorca, Spain}
\author{C. Anteneodo} 
\affiliation{Department of Physics, PUC-Rio, Rua Marqu\^es de S\~ao Vicente, 225, 22451-900, Rio de Janeiro, Brazil}
\affiliation{Institute of Science and Technology for Complex Systems, Rio de Janeiro, Brazil}

\begin{abstract}

Internal feedbacks are commonly present in biological populations and can play a crucial role in the emergence of collective behavior. 
We consider a generalization of Fisher-KPP equation to describe the temporal evolution of the 
distribution of a single-species population. This equation includes the elementary processes of random motion, reproduction and, importantly, nonlocal interspecific competition, which introduces a spatial scale of interaction. 
Furthermore,  we take into account feedback mechanisms in diffusion and  growth processes, mimicked through density-dependencies controlled by exponents $\nu$ and $\mu$, respectively. 
These feedbacks include, for instance, anomalous diffusion, reaction to overcrowding or to rarefaction of the population, as well as Allee-like effects.  
We report that, depending on the dynamics in place, the population can self-organize 
splitting into disconnected sub-populations, in the absence of environment constraints. 
Through  extensive numerical simulations,  we investigate the temporal evolution 
and stationary features of the population distribution in the one-dimensional case. 
We discuss the crucial role that density-dependency has on pattern formation, particularly on fragmentation,   which can bring important consequences 
to processes such as epidemic spread and speciation.

\end{abstract}

\maketitle

\section{Introduction}

Population fragmentation is characterized by critical changes in the spatial distribution of individuals, creating isolated sub-groups of a given initial population. 
This phenomenon has important consequences for secondary processes such as epidemic spreading, 
species invasion~\cite{Ludwig1979} or also speciation~\cite{baptestini2013}. 
Fragmentation is often attributed to landscape heterogeneity which embraces the spatial distribution of geographical and environmental features~\cite{turner2001}.
If natural barriers are sustained for long periods of time, fragmentation can be induced~\cite{baptestini2013}.

This scenario has been vastly studied in the context of metapopulation theory, which takes into account the ecological landscape heterogeneity~\cite{hanskiBook}. 
 The degree of fragmentation of the landscape, which is imposed to the population, is well-known to play an important role, determining the species richness and 
ecosystem stability against external perturbations~\cite{levin,hanskiBook,diversity}. 
But, regardless of  environment heterogeneity, arrangements of 
individuals in space can also emerge solely from their interactions, 
 bringing critical consequences to the evolutionary dynamics and social behavior of living organisms 
(Refs.~\cite{ecoevo1,ecoevo2,ecoevo3,ecoevo4,Pigolotti2012}).

Precisely, we explore in this work under which conditions population dynamics can self-induce fragmentation in the absence of external barriers. 
A previous study has pointed out that  spatial patterns in the population distribution can become disconnected  when individuals' dispersal is subdiffusive~\cite{Colombo2012}. We extend this investigation, delving in the characterization of the fragmentation process and assuming a more general nonlinear dynamics, where besides dispersal also growth can be regulated by population concentration.
Namely, we generalize the well-known Fisher-KPP equation \cite{fisher1937,kolmogorov1937}, which includes standard diffusion and logistic growth of the population, by means of power-law density-dependencies in 
the rates of those processes.

Density-dependent mobility can arise due to the environment structure~\cite{bacporous,porous}, 
but it can also originate from  complex biological and social reactions, in response to 
 overcrowding or rarefaction of the population density~\cite{Cates2010,murray2002,cristobal2006,Kenkre2008,Colombo2012,celia2001,celia2005,celia2007}. 
 For instance, in populations of insects, it has been observed that the diffusion coefficient can be enhanced or harmed by population concentration~\cite{murray2002}. In this and many other examples~\cite{murray2002,newman,gurtin,kareiva,nldiffGene}, a power-law form for the diffusion coefficient was used as phenomenological description.

Population growth can also be governed by density-dependent factors~\cite{nldiffGene,dos2014generalized,cabella2011data,dos2015models,cabella2012effective, martinez2009generalized,martinez2008continuous}. 
For instance, related to the Allee effect~\cite{allee},  
the per capita reproduction rate vanishes in the low concentration limit. 
But, there are also cases where  reproduction is favored when the concentration is low, 
due to the absence of overpopulation disadvantages~\cite{parasite,Colombo2018}.

On top of all that, our model considers resource sharing within a given spatial range, through a nonlocal competition term.
In vegetation, for instance, long roots can induce water competition at distance~\cite{escaff2015,tarnita2017,fernandez2019front}. 
The release of toxic substances in the environment can also promote death in spatial scales much larger than individual's size~\cite{Fuentes2003,calleja2007relationship}. 
Such mechanisms generate an effective kernel, also known as influence function, that introduces a distance dependent spatial coupling~\cite{martinez2014minimal}. Under some conditions, this spatial coupling can promote spatial instability, a key ingredient for pattern formation~\cite{emilio,martinez2014minimal,Pigolotti2007}.

It is worth noting that our modeling based on the Fisher-KPP equation 
aims to describe the temporal evolution of population distributions, but also of gene distributions, niche occupation or traits~\cite{fisher1937}. Then, the fragmentation process that we focus in this work has an interesting ambiguity, which can be translated onto speciation, for instance~\cite{speciation,Pigolotti2007}.

The paper is organized as follows.
In section \ref{sec:model}, we define the generalization of the Fisher-KPP equation that we use as paradigmatic model. In section \ref{sec:linear}, we obtain analytical results to define the conditions for pattern formation and in Sec. \ref{sec:num}, we present the main results from numerical simulations, aiming to characterize the different classes of patterns, particularly fragmented ones.
 In Sec. \ref{sec:sum} a summary and discussion of the main results and possible extensions are presented.

\section{Model}
\label{sec:model}

 We consider the following generalization of the one-dimensional 
Fisher-KPP equation~\cite{fisher1937} 
for the spatial distribution of  one-species populations
 
\begin{equation}
\partial_t \rho(x,t) = \partial_{x}(D(\rho) \partial_x \rho) +  f(\rho)\rho - b\rho\int_{-\infty}^{\infty} \gamma(x,y)\rho(y) dy\, .
\label{premaineq}
\end{equation}
The first term on the right-hand side of  Eq.~(\ref{premaineq}) 
corresponds to nonlinear diffusion, 
where the diffusion coefficient $D(\rho)$ depends on the local density $\rho(x,t)$. 
The second term regulates  reproduction, which occurs with growth rate per capita $f(\rho)$,  
that also depends on the local density.  
The last term represents the nonlocal intraspecific competition, where $b>0$, and the influence function $\gamma$ 
sets how the interaction depends on the distance.

Following the motivations given in the Introduction,  we investigate the class of dynamics where diffusion and growth coefficients have power-law density-dependencies, namely, 
\begin{align}
D(\rho) &= D\rho^{\nu-1},\\
f(\rho) &= a\rho^{\mu-1},
\end{align}
where $D$, $a$, $\nu$ and $\mu$ are positive parameters.
For logistic effect (referring to limited resources), we must have $\mu<2$, to ensure that the population size  remains  bounded.

Before proceeding, we nondimensionalize Eq.~(\ref{premaineq}), by defining the scaled variables
\begin{align}  \nonumber 
\rho^\prime &= \rho/\rho_0 , \\ \nonumber 
t^\prime    &= a \rho_0^{ \mu-1}t,\\ \label{scalings3}
x^\prime    &=  \sqrt{a\rho_0^{\mu-\nu}/D} \;x ,
\end{align}
where $\rho_0 = (b/a)^{1/(\mu-2)}$ is the uniform stationary solution, that  
becomes $\rho_0^\prime =1$. 
Then, substituting the scaling relations (\ref{scalings3}) into Eq.~(\ref{premaineq}) 
and eliminating the prime superindexes, Eq.~(\ref{premaineq}) becomes
\begin{equation} \label{maineq}
\partial_t \rho(x,t) = \partial_{x}(\rho^{\nu-1} \partial_x \rho) +  \rho^{\mu} -  \rho\int_{-\infty}^{\infty} \gamma(x-y) \rho(y) dy\, .
\end{equation}
In this way, the exponents $\mu$ and $\nu$ are the only remaining parameters, once fixed kernel $\gamma$.

\section{Linear stability analysis}
\label{sec:linear}

Following the standard procedure, we assume a small perturbation around the nontrivial homogeneous steady state, i.e., $\rho(x,t) = 1 + \varepsilon(x,t)$.

Linearization of Eq.~(\ref{maineq}) yields
\begin{equation}
\partial_t \varepsilon =  \partial_{xx} \varepsilon + (\mu-1)  \varepsilon -  \int_{-\infty}^{\infty} \gamma(x-y) \varepsilon dy ,
\end{equation}
which in Fourier space becomes
\begin{equation}
\partial_t \tilde\varepsilon(k,t) = \lambda(k)\tilde\varepsilon(k,t)\, ,
\end{equation}
where  the tilde mark indicates Fourier transform, and the rate $\lambda(k)$  is 
given by the dispersion relation
\begin{equation}
\lambda(k) = -   k^2  -  \tilde\gamma(k) +  \mu-1   \, .
\label{dispersion}
\end{equation}

Pattern formation occurs when there is a certain dominant mode $k^\star$ that stands out in the dispersion relation, that is, yielding maximum positive rate $\lambda(k^\star)$~\cite{Cross1993}. 
The condition for pattern formation ($\lambda(k^\star)>0$) depends on the profile 
of the influence function $\gamma$ that must introduce a well-defined spatial scale of interaction~\cite{Pigolotti2007}.
The simplest form that verifies this property, promoting spatial instability, is the homogeneous influence function, which is constant inside a certain region of width $2\ell$,

\begin{equation}
\gamma(x-y) = \frac{1}{2\ell} \Theta (\ell - |x-y|) \, ,
\label{kernel}
\end{equation}
being non-null only if $|x-y|<\ell$.
Therefore, its Fourier transform is 

\begin{equation}
\tilde{\gamma}(k)=\sin(k\ell)/(k\ell).  
\label{gamatilde}
\end{equation}

The first term in Eq.~(\ref{dispersion}), associated to diffusion,
is always negative,  tending to stabilize the homogeneous state. 
The term $\tilde\gamma(k)$ given by Eq.~(\ref{gamatilde}), associated with nonlocality,  
takes positive and negative values and therefore can contribute to 
destabilize the homogeneous state, therefore, giving rise to pattern formation. 
Additionally, the nonlinearity $\mu\neq1$ shifts the dispersion relation 
with respect to the linear case ($\mu=1$), 
contributing to destabilization when $\mu>1$ and to stabilization when $\mu<1$. 
Notice that the diffusion exponent $\nu$ does not appear 
explicitly in the dispersion relation. 

The dominant mode $k^\star$, which is the maximum of $\lambda(k)$, 
can be approximated by $ k^\star \ell \simeq 3\pi/ 2 $~\cite{Colombo2012}. 
Its rate of exponential change is positive if
\begin{equation}
\mu > \mu_p \equiv  (k^\star)^2   - \frac{1}{k^\star\ell} +1. 
\label{eq:mup}
\end{equation}
This constitutes the frontier for the onset of patterns. 
Moreover, when patterns appear, the number $m$ of peaks can be estimated by
\begin{equation}
m= \frac{k^\star L}{2\pi}  \simeq  0.715\frac{L}{\ell},
\label{eq:maxima}
\end{equation}
where $L$ is the system size.

Notice that nonlinearities are also contained in the spatial and time scales, 
according to Eqs. (\ref{scalings3}), hence they influence pattern wavelength 
and growth rate. 
Therefore, although $\nu$ does not appear explicitly in Eq.~(\ref{dispersion}),  
it has an indirect influence.

\section{Numerical results}
\label{sec:num}

\begin{figure}[b!]
\includegraphics[width=\columnwidth]{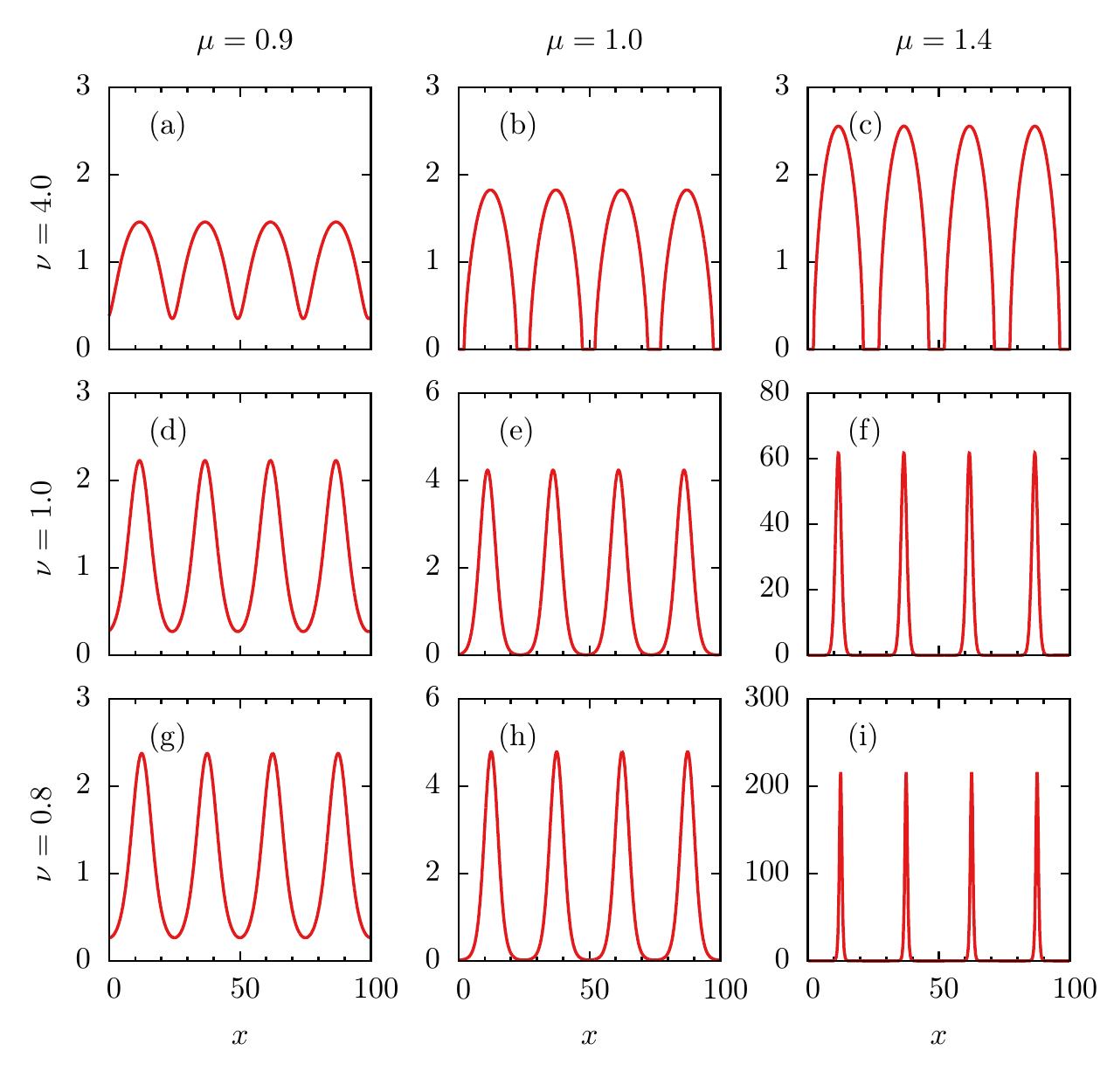}
\caption{Shape of stationary patterns $\rho(x)$ for different values of $\nu$: 
$0.8$ (superdiffusion),   $1.0$ (normal diffusion) and $4.0$ (subdiffusion), 
and different values of the growth exponent $\mu$: 0.9, 1.0 and 1.4.}
\label{fig:patterns}
\end{figure}

Numerical integration of Eq.~(\ref{maineq}) was performed using a forward-time, 
centered-space scheme~\cite{numericalrecipes}, 
considering a one-dimensional domain with periodic boundary conditions. 
Starting from the homogeneous steady state $\rho_0=1$, 
with the addition of a white-noise perturbation, uniform 
in $[-\delta \rho_0,\delta \rho_0]$ with $\delta \rho_0=10^{-2}$, 
we let the dynamics evolve during a time long enough for the stationary regime to be achieved.

In all the numerical simulations, we fixed the system size $L=100$ and 
the competition interaction range $\ell=20$.
As a consequence of this choice, Eq.~(\ref{eq:maxima}) predicts that, when there are 
patterns ($\mu>\mu_p \simeq  0.84$ in this case), the expected number $m$ of peaks is $m=3.75$, within the linear approximation. 
Therefore, more likely we observe 4 peaks.

Typical pattern shapes that emerge in our numerical simulations are presented in Fig.~\ref{fig:patterns} 
(see also Appendix for further details), 
 for different values of $\nu$ and $\mu$ in the region where $\lambda(k)>0$. 
In the standard case $\mu=\nu=1$, each individual peak has a Gaussian shape. 
  But when feedbacks are taken into account, mobility and reproduction 
	rates respond to degree of agglomeration of 
	individuals. Then,  peaks tend to be more platykurtic (leptokurtic) when 
$\nu>1$ ($\nu<1$), since the diffusion rate vanishes (diverges) at low densities.
With respect to  exponent $\mu$, it is evident that the patterns that emerge when $\mu <1$ have 
a minimal value which is noticeably different from zero, in contrast to the cases $\mu \ge 1$. 
These features can be associated to the type of density-dependent feedback (ruled by $\mu$): 
when $\mu<1$ growth is enhanced in low density regions, rising the level in between clusters;
while for $\mu>1$, the opposite effect occurs.  
The combination of  diffusion and growth nonlinearities generates the diverse profiles shown in Fig.~\ref{fig:patterns}.
Next, we will discuss how these different profiles can emerge, focusing on the characterization and definition of fragmented states (Figs.~\ref{fig:patterns}b-c).

\begin{figure}[h!]
\includegraphics[width=\linewidth]{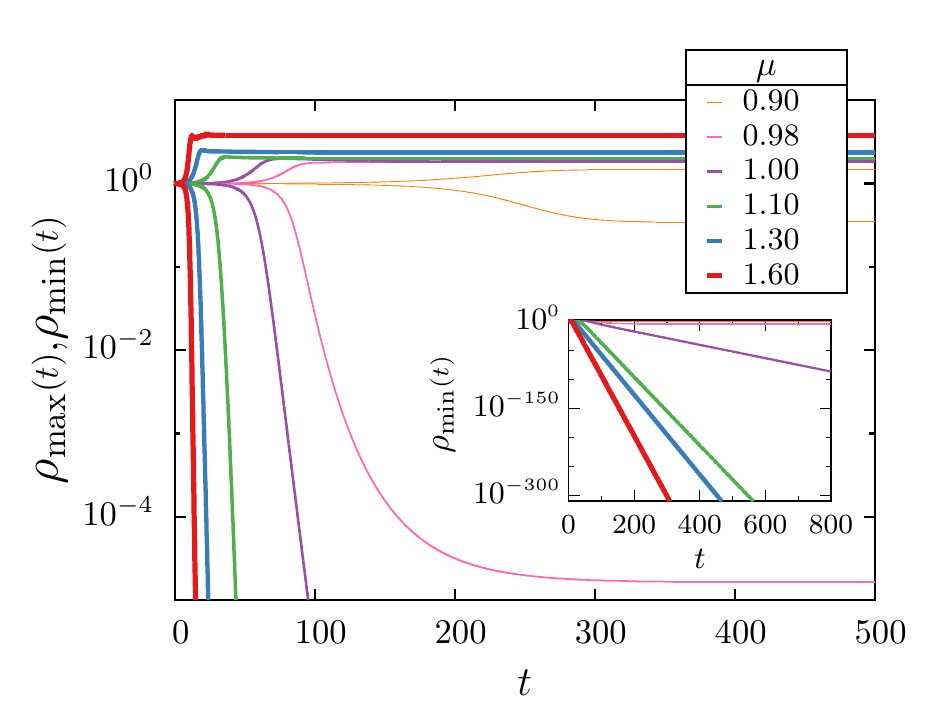}
\caption{Temporal evolution of the maximal and minimal densities
$\rho_{\text{max}}(t)$ and $\rho_{\text{min}}(t)$, for $\nu=4.0$ and 
values of $\mu$ indicated in the legend 
(thicker lines correspond to larger values of $\mu$). Inset: $\rho_{\text{min}}(t)$ on a larger scale.
}
\label{fig:evolution}
\end{figure}

In order to identify  the  fragmentation process, 
we followed the temporal evolution of the  lowest value of the concentration of individuals,  
$\rho_{\text{min}}(t)$. 
Representative cases are shown in Fig.~\ref{fig:evolution}, where besides
the minimal value, also the maximal one $\rho_{\text{max}}(t)$ is plotted.

\begin{figure}[h!]
\includegraphics[width=1.0\columnwidth]{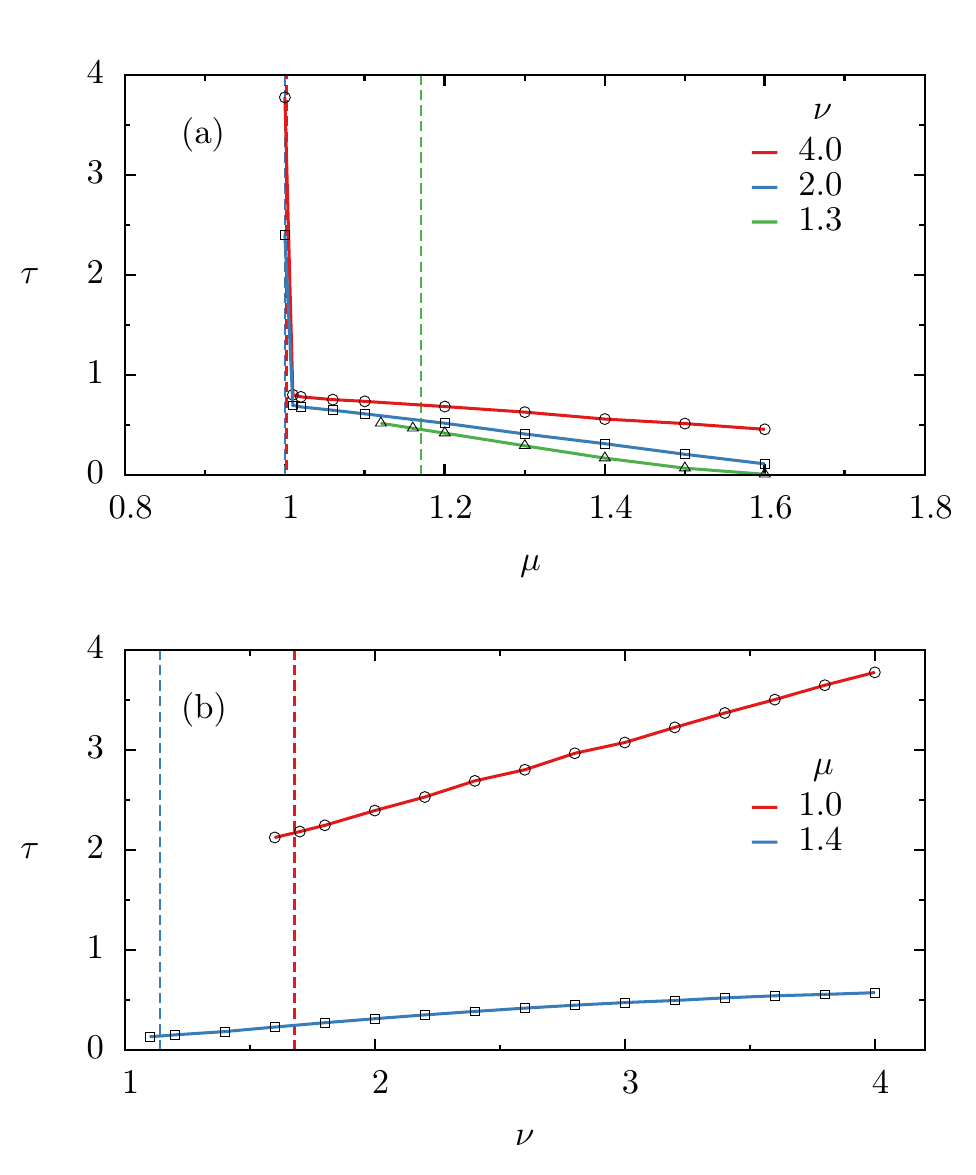}
\caption{Characteristic time $\tau$ of the exponential decay of $\rho_\text{min}(t)\sim \exp(-t/\tau)$ 
as a function of:  (a) $\mu$ (for fixed values of $\nu$) and (b) $\nu$ (for fixed $\mu$), as indicated in the figures. The dashed vertical lines correspond to the values at which fragmentation occurs as explained in the text.
}
\label{fig:time}
\end{figure}

We observe that for enough small values of $\mu$, $\rho_{\text{min}}(t)$ 
stabilizes in a finite level. 
In contrast for $\mu$  larger than a critical value ($\mu_c=1$, in the case of Fig.~\ref{fig:evolution}),  
$\rho_{\text{min}}(t)\sim \exp(-t/\tau)$, decreasing  exponentially with time down to the computational limit ($\varrho\sim 10^{-320}$). 
Values of the characteristic time $\tau$ are shown in Fig.~\ref{fig:time}, 
for different values of the exponents, 
including the cases shown in Fig.~\ref{fig:evolution}.

\begin{figure}[h!]
\includegraphics[width=1.0\columnwidth]{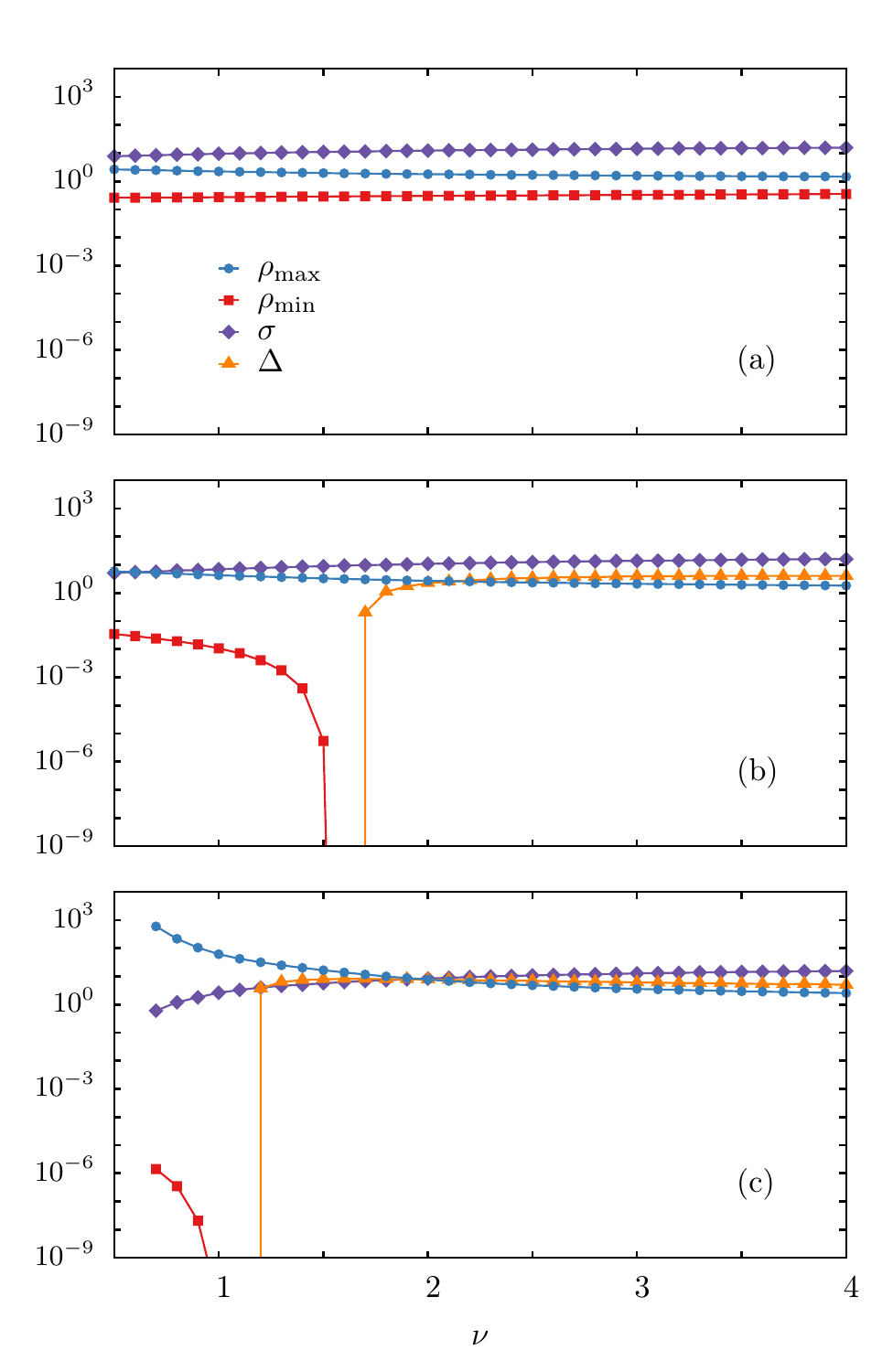}
\caption{Stationary values of maximal density ($\rho_{\text{max}}$), 
minimal density ($\rho_{\text{min}}$), width at half height  ($\sigma$) 
and valley width  ($\Delta$) as a function of $\nu$, for
(a) $\mu=0.9$, 
(b) $\mu=1.0$, 
(c) $\mu=1.4$. 
}
\label{fig:var_nu}
\end{figure}

The numerical outcomes suggest the emergence of disconnected clusters,
separated by non-populated regions, when $\nu$ and $\mu$ obey certain conditions.
In order to further characterize the fragmented patterns 
and their emergence conditions, 
besides the stationary values ($\rho_{\text{max}}$ and $\rho_{\text{min}}$), 
the width $\sigma$ of each cluster at half height  
and the length $\Delta$ of the region where $\rho$ attains $\varrho$, which we consider as null density~\footnote{We identify this region as null density since, within it, the null state is numerically stable. Specifically, setting to zero the density values they remain stable.}. 
Results are shown in Fig.~\ref{fig:var_nu}, 
varying diffusion exponent $\nu$ while keeping the growth exponent $\mu$ constant.
For $\mu=0.9$ (Fig.~\ref{fig:var_nu}a), the shape of the patterns is almost insensitive to 
$\nu$. Importantly, we do not detect a region where the density vanishes, for this reason 
values of $\Delta$ do not appear in the plot. As a consequence, fragmentation does not exist.
Differently, in Fig.~\ref{fig:var_nu}b-c,  
a sharp drop of $\rho_{\text{min}}$ is observed as $\mu$ increases. 
Concomitantly,  a non null $\Delta$ is detectable in these cases.
Then, we identify that, beyond a critical value of $\nu$ (that decreases with $\mu$)
patterns become fragmented.

\begin{figure}[h!]
\includegraphics[width=1.0\columnwidth]{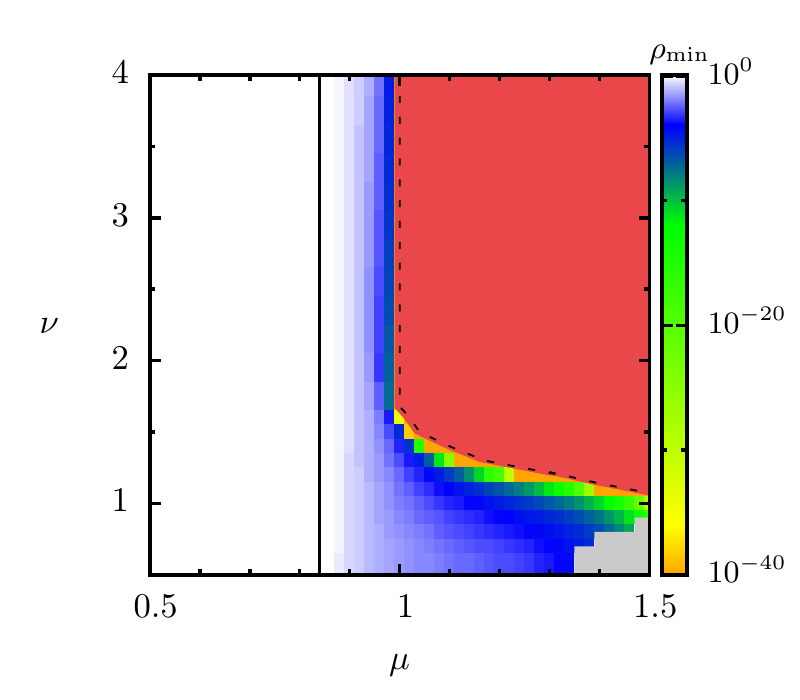}
\caption{Phase diagram, in the plane $\mu-\nu$. 
The color scale represents the stationary minimal density $\rho_{\text{min}}$.%
The vertical solid line at $\mu_p \simeq 0.84$ delimits upperly the domain where no patterns are formed, 
according to Eq.~(\ref{eq:mup}). In that region, $\rho_{\text{min}}=\rho_{\text{max}}=1$. 
Above $\mu_p$ patterns emerge, whose minimal value gradually decreases. 
The dashed line separates the non-fragmented region from the fragmented one (red). 
The gray region corresponds to values not calculated due to computational limitations.
}
\label{fig:df}
\end{figure}

A picture of the regions in the plane $\mu-\nu$ where patterns develop, and 
where they fragment or not, is presented in Fig.~\ref{fig:df}, obtained from numerical simulations.  
The white region at the left of the vertical solid line corresponds to values 
of the exponents for which no patterns arise, in agreement with condition Eq.~(\ref{eq:mup}),
while patterns emerge in the complementary domain.
The solid (red) area denotes patterns that are fragmented, in the sense defined above.

Fragmentation occurs depending on a balance between diffusion and growth at low densities. 
Looking at Fig.~\ref{fig:df}, we see that fragmentation is favored when diffusion coefficient and per capita reproduction increase superlinearly with population concentration ($\nu$ and $\mu$ larger than one).
Differently, when $\nu$ and $\mu$ are small, diffusion and growth per capita diverge at low densities, promoting fast occupation of non-populated regions, hence connecting clusters.

\begin{figure}[h!]
\includegraphics[width=1.0\columnwidth]{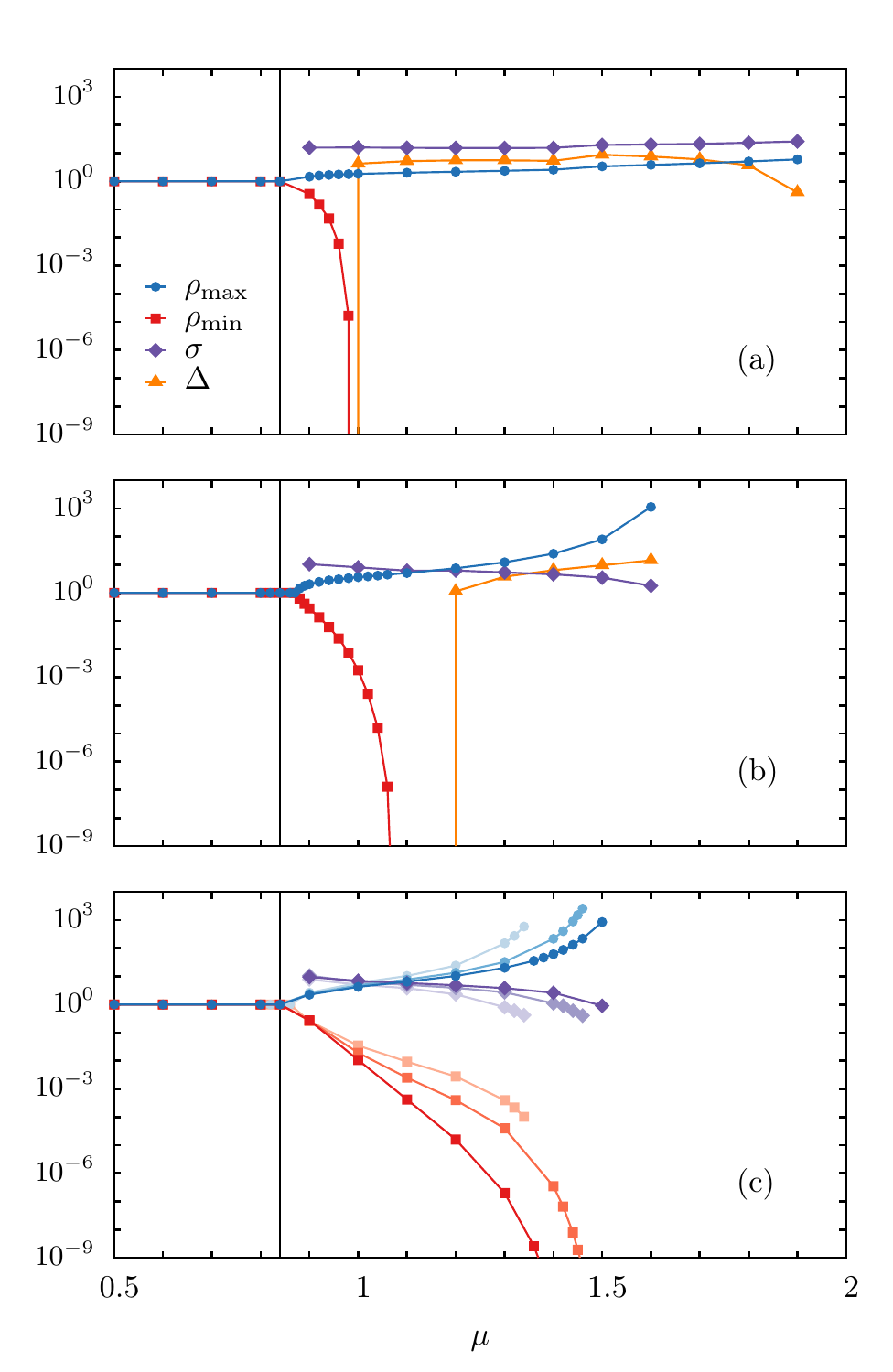}
\caption{Stationary values of the maximal density ($\rho_{\text{max}}$), 
minimal density ($\rho_{\text{min}}$), width at half height  ($\sigma$) 
and valley width  ($\Delta$) as a function of $\mu$, for
(a) $\nu=4.0$, 
(b) $\nu=1.3$ and
(c) $\nu=0.5$, 0.8 and 1.0, where more intense colors correspond to higher values of $\nu$.
The vertical lines represent $\mu_p \simeq 0.84$.
}
\label{fig:var_mu}
\end{figure}

More details about the pattern shape transitions are shown in Fig.~\ref{fig:var_mu}.
We see that beyond the critical frontier of fragmentation, 
when $\nu=4$ (Fig. \ref{fig:var_mu}a), there is a smooth variation 
in the shape quantities $\sigma$, $\Delta$ and $\rho_\text{max}$,   as in the cases of Fig.\ref{fig:var_nu}. 
(Except that as $\mu \to 2$, nonlinearities affect the amount of peaks $m$ 
and hence the measured quantities.)
But when $\nu$ becomes small, the behavior of pattern features changes. 
In Fig.~\ref{fig:var_mu}b-c, we notice a region where the shape quantities vary exponentially 
with $\mu$, followed by a regime where changes occur faster. 
Note for instance that, while the height of a peak $\rho_\text{max}$ rapidly increases, its
width $\sigma$ decreases with $\mu$, 
features that suggest that each peak tends  to approach a Dirac delta-like profile. 
The effect is accentuated for small $\nu$, as can be seen in 
Fig.~\ref{fig:var_mu}c. 

This behavior brings numerical difficulties, that prevent determining 
whether   a Dirac delta is attained or not  for finite $\mu$, since 
the increments  $dx$ and $dt$ used in simulations must be reduced, 
hence increasing the computational cost.  
It is worth to remark that despite the dependency of $\rho_\text{min}$ 
with the model exponents is similar to those in Fig.~\ref{fig:var_nu}b-c, 
mainly the sharp drop feature, we could not follow the behavior 
until $\varrho$ is attained (or not) due to strong instability 
in numerical integration when $\mu\to 2$ (see gray region in Fig.~\ref{fig:df}). 
Such complications compromises a definite conclusion regarding the fragmentation process for large values of $\mu$, specially for small $\nu$. 
Particularly when $\nu\leq 1$ (see Fig.~\ref{fig:var_mu}c), 
despite there is some indication that 
there exists a critical value of $\mu$ for which fragmentation occurs, 
we could not observe the complete abrupt drop of the minimum value $\rho_{\text{min}}$ to zero. 

Finally, concerning the temporal aspects of the pattern shape transition, 
we address further comments related to Fig.~\ref{fig:time}. 
For large values of $\nu$ ($\nu=2.0,4.0$ in Fig.~\ref{fig:time}a), 
the time $\tau$ increases as $\mu$ decreases, exploding at the critical value. 
In these cases, the relaxation time towards zero (when fragmentation occurs) 
and the relaxation towards a finite minimum population {(otherwise), 
suffer a drastic change. 
That is, together with the transition  of the minimum value of the 
stationary density $\rho_{\text{min}}$, there is a transition in 
the dynamics timescale, which becomes slower when $\mu$ increases 
(see Fig.~\ref{fig:time}a-b).
In contrast, there are other cases where a drastic change in the timescale 
is not observed, and there is continuity of the values of $\tau$ 
 across the fragmentation boundary. 
That is, the decay time towards a finite level (at the left of 
the vertical lines in the figure) or 
towards zero (at the right of the vertical lines) does not suffer 
a discontinuity.  
This indicates that depending on the region of the $\mu-\nu$ plane, 
the transition to fragmentation can occur in two distinct ways.

The relation between nonlinearities and patterns shape and its implication 
for population dynamics will be discussed in the following Section.

\section{Summary and Discussion}
\label{sec:sum}

Using as starting point a nonlocal Fisher-KPP equation, which became a relevant description 
in mathematical biology~\cite{Fuentes2003,emilio,clerc,martinez2014minimal,tarnita2017,DaCunha2011}, we introduce density-dependent feedbacks in diffusion 
and growth processes and investigate their effects in shaping the population distribution. 
We choose the particular form of power-law dependencies on the density, 
that allow to contemplate a large class of responses to  population density, as  found in populations of insects, bacteria, vegetation, among other cases, where diffusion and growth can be either enhanced or harmed by the concentration of individuals.

The emerging patterns have shapes ranging from mild oscillations around a reference level  
to disconnected clusters. 
The growth regulatory mechanisms represented by $\mu$ 
are crucial for the emergence of patterns as well as for fragmentation.
The same can be said about the type of diffusion controlled by $\nu$, 
despite diffusion has in general homogenizing effects.

Since population dynamics equation, Eq.~(\ref{maineq}),  
is nonlinear and nonlocal, our main results, beyond linear stability analysis, 
were obtained through numerical simulations. 
Insights can also be brought from related models with power-law dependencies, 
although they do not contain nonlocality, and boundary conditions are not periodic 
~\cite{Colombo2018,porous,bukman,plastino}. 
In these works, peaks similar to those found in the present context, 
 ranging from concave to sharp peaks, were observed. 
In some cases the solutions fall  into the class of a generalized 
Gaussian shape~\cite{porous,bukman,plastino,celia2005}. 
This motivated us to propose a periodic extension of that ansatz 
for the profiles shown in Fig.~\ref{fig:patterns}, 
namely  Eq.~(\ref{eq:extension}),  which describes remarkably well the numerical 
patterns (see Fig.~\ref{fig:exp_qgauss} in the Appendix). 
Parameter $\beta$ in Eq.~(\ref{eq:qgauss}) can be used to characterize pattern shape.
Notice that $\beta=0$ corresponds to a Gaussian, $\beta>0$ ($<0$)
to platykurtic (leptokurtic) clusters. 
In particular, for $\beta>0$, an individual cluster have compact-support property. 
We found that such kind of profiles is associated to the emergence of fragmentation, 
with the additional condition of non-overlap, $2x_0 < \Lambda$, as defined in the Appendix. 
These two conditions reproduce well the fragmented-patterns 
region in the phase diagram (Fig.~\ref{fig:df}), where 
$\rho_{\text{min}}\to \varrho$.

Particularly, we focused on the self-induced population fragmentation, 
determining the conditions that non-linearities must obey. 
Briefly, we observed that fragmentation is favored when growth and diffusion coefficients 
are positively correlated with population density. Moreover, it arises from a complex 
interplay between growth and dispersal processes (see critical line in Fig.~\ref{fig:df}).

Regarding the definition of fragmentation, 
previous models for pattern formation, that helped to explain self-organization 
in mussels~\cite{koppel2008}, bacteria~\cite{ben1994}, vegetation 
under the sea~\cite{ruiz2017} and in semi-arid ecosystems~\cite{tarnita2017,escaff2015}, 
produce an arrangement of high density clusters interleaved by low density regions. 
In some cases, when clusters are sharply defined or well spaced, the population level in between can be very low. 
More specifically, in these cases, population concentration is expected 
to decay exponentially as we move away from the peaks 
(see for instance Ref.~\cite{escaff2015}). 
Taking into account that a biological population is constituted by a finite number of individuals, the occurrence of 
very low densities in the mean-field description can be associated with an effective fragmentation of the population. 
This is because, in the continuous density description, it is possible to emulate the finiteness of the population by means of a threshold value,  inversely proportional to the number of individuals and below which the density is considered null.  
Under this perspective, the region for fragmentation in the phase diagram of Fig.~\ref{fig:df} 
would be effectively enlarged as the number of individuals diminishes. 
In contrast, according to our model density-dependent feedbacks drive the population density between clusters to zero in the long-time limit, such that the stationary profiles are composed by clusters with the compact-support property. As a consequence, 
actual fragmentation occurs and it is robust independently of the number of individuals (i.e., the threshold value) considered.

Beyond the nonlocal interactions embodied in the influence function, 
when there are isolated clusters, individuals are only in direct contact 
with those within the same cluster. 
This restrains the propagation of contact processes, 
such as diseases or information, that are transfered from one individual to another. 
Initiating the contagion inside one isolated cluster, 
the affected population would be confined, while, 
in non-fragmented patterns, information would 
percolate to the whole population.  
In fact, arrangements that emerge solely from the interactions, 
were shown to bring critical consequences to populations dynamics 
~\cite{ecoevo1,ecoevo2,ecoevo3,ecoevo4,Pigolotti2012}.
Furthermore, as vastly studied, 
fragmented habitats play an important role in the stability 
and diversity of ecosystems~\cite{levin,diversity}.
%
%
In our case, the distinct profiles which emerge from the dynamics are also expected to 
influence population fate. Therefore, as a perspective of future work, it may be worth to study the 
coevolution of contact processes and population dynamics ruled by Eq.~(\ref{maineq}).

Lastly, it is important to have in mind that, in Nature, fragmentation may arise not solely from 
either the heterogeneity of the environment or the selforganization 
of the population but from the interplay between both features, that are interdependent, 
reciprocally influencing each other.
In this respect, it would be interesting to  investigate in future works  
their reciprocal influence.

\appendix

\section*{Appendix: Shape of patterns}
\label{app}
\renewcommand{\theequation}{A.\arabic{equation}}

\begin{figure}[h!]
\includegraphics[width=\columnwidth]{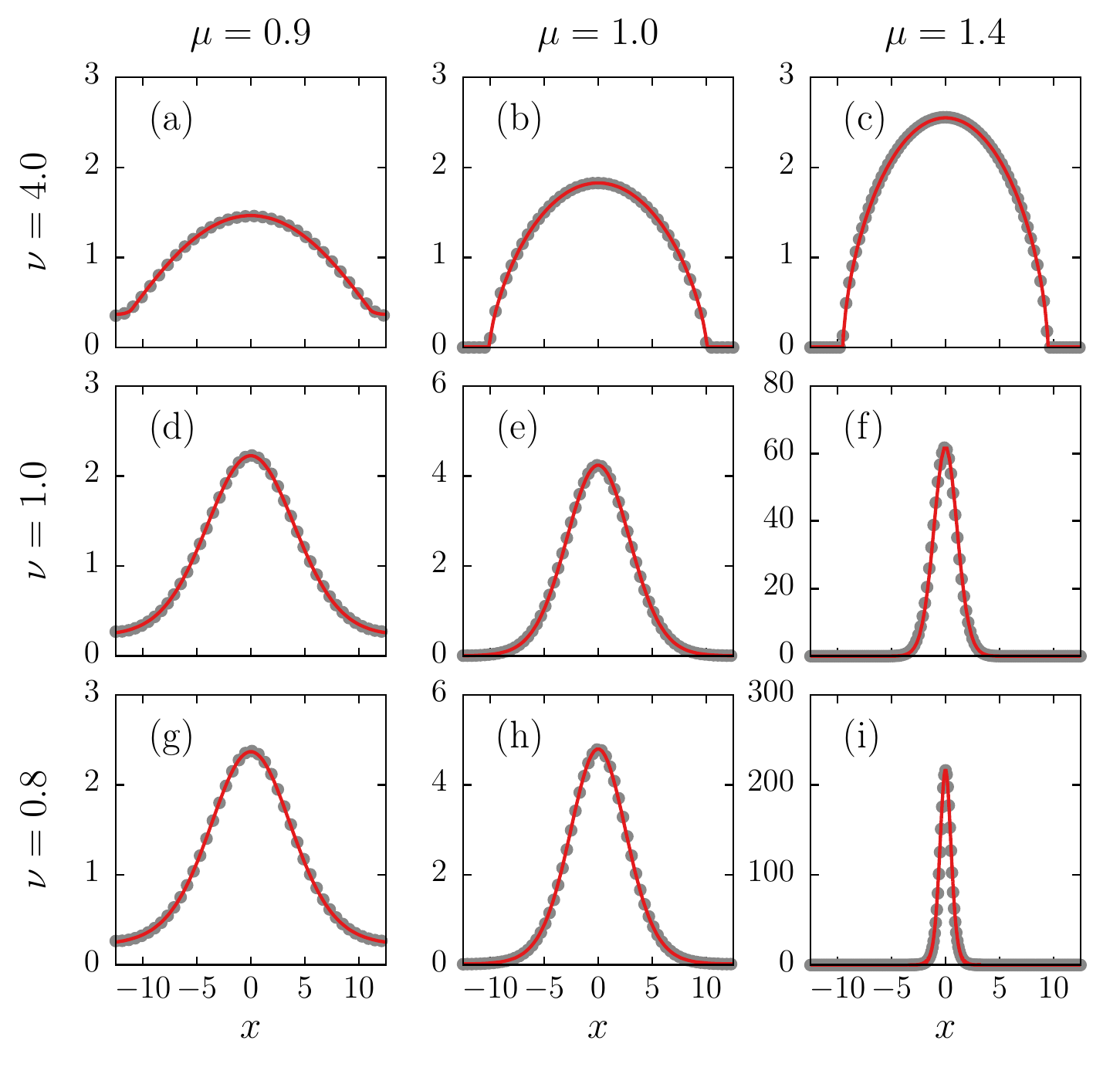}
\caption{Stationary profiles obtained numerically together with 
the description given by  Eq.~(\ref{eq:qgauss}), 
for the values of $\mu$, $\nu$ indicated on the figure.  
Fitting values of the parameters are given in Table~\ref{table:qgauss}. 
}
\label{fig:exp_qgauss}
\end{figure}

We show in this section that the patterns that emerge from the generalized Fisher-KPP Eq.~(\ref{maineq}) can be described in very good approximation by a the periodic extension of a generalization of the Gaussian function. 
Inspired by the form of the solutions of the (nonlinear diffusion) porous media equation~\cite{porous} and other related ones~\cite{celia2001,celia2005,celia2007,bukman,plastino},
we consider the ansatz
\begin{equation}
f(x)=A\left(1-\frac{\beta x^2}{2s^2}\right)_+^{1/\beta},
\label{eq:qgauss}
\end{equation}
where $A$, $s$ are positive constants, and $\beta$ is real. 
The subindex ``+'' indicates that the expression between parentheses must 
be zero if its argument is negative. In this case the function vanishes 
outside the interval $[-x_0,x_0]$, with $x_0 =  s\sqrt{2/\beta}$.

\renewcommand{\arraystretch}{1.2}
\begin{table}[t!]
\footnotesize
\begin{tabular}{|c|c|r|r|r|}
\hline
  	        &  		     & $\mu=0.9$ 			  & $\mu=1.0$ 			& $\mu=1.4$ 		\\ 
\hline 
						& $A$  	   &	1.4645[7] 			&  1.8264[7] 			  &  2.5578[6]  		\\ 
$\nu=4.0$ 	&$\beta$   &  0.793[3]   			&  1.428[3]    			&  1.594[2]  	\\ 
			      & $s$      &  8.761[9]   			&  8.521[7]  				&  8.501[4] 		\\ 
 \hline 
						& $A$      &  2.222[1]  		  &  4.245[2]  				&  61.5[2]  		\\ 
$\nu=1.0$ 	& $\beta$  & -0.316[3]     		&  -0.076[2]  			&  -0.11[1] 	\\ 
						& $s$      &  4.149[6]  			&  2.912[3] 				&  1.077[6] 		\\ 
 	\hline 
						& $A$ 		 &  2.367[2]  			& 4.797[8] 					&  213.0[4]		  \\ 
$\nu=0.8$ 	& $\beta$  & -0.395[4]				& -0.153[7] 				& -0.259[8] \\ 
						& $s$  		 &  3.777[9]  			& 2.560[9]					&  0.472[2]	  \\ 
\hline 
\end{tabular}
\caption{Parameter values from the (nonlinear least-square) 
fitting of Eq.~(\ref{eq:extension}), in the interval 
$[-\Lambda/2,\Lambda/2]$ (with $\Lambda=50$), to stationary patterns displayed 
in Fig.~\ref{fig:exp_qgauss}, after centering a maximum at $x=0$. 
The square brackets contain the estimated error in the least significant figure 
(e.g., the notation 213.0[4] stands for $213.0 \pm 0.4$). 
}
\label{table:qgauss}
\end{table}

If $\beta \to 0$,   Eq.~(\ref{eq:qgauss}) recovers the Gaussian function, 
otherwise, this function represents the generalized Gaussian that 
arises within Tsallis statistics~\cite{tsallis2009introduction}.  

 To describe the steady states observed in our case, we consider 
the periodic extension of Eq.~(\ref{eq:qgauss}) with period $\Lambda$, 
that is

\begin{equation}
f^\text{ext}(x) = \sum_{k\in \mathbb{Z}}   f(x-k\Lambda).
\label{eq:extension}
\end{equation}

Figure~\ref{fig:exp_qgauss} shows  stationary patterns adjusted 
by Eq.~(\ref{eq:qgauss}) and the Table~\ref{table:qgauss} 
shows the fitting parameters. 
Only one wavelength $\Lambda$ of $\rho(x)$ 
(between successive minima of $\rho$) is represented.

\begin{figure}[h!]
\includegraphics[width=\columnwidth]{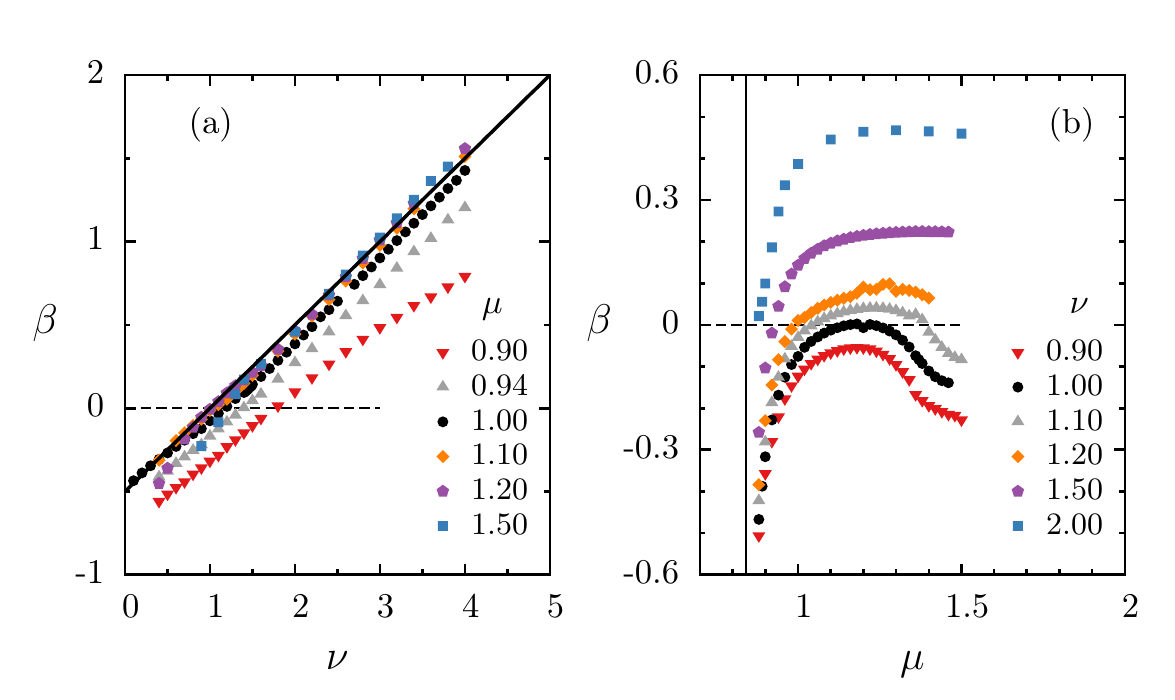}
\caption{Ansatz exponent $\beta$ as a function 
of model exponents $\nu$ (a), $\mu$ (b). 
The solid line in (a) corresponds to $\beta=(\nu-1)/2$, drawn for comparison.
The vertical solid line in (b)  represents $\mu=\mu_p$. 
The value $\beta=0$ is highlighted by dashed horizontal lines.
}
\label{fig:beta}
\end{figure} 

We observe, in Fig.~\ref{fig:exp_qgauss} and Table~\ref{table:qgauss}, 
that when $\nu=\mu=1$, 
the shape is nearly Gaussian, since $\beta\simeq 0$.  
Gaussian approximations were found for a similar evolution equation with 
normal diffusion \cite{Barbosa2017}.
But when the exponents become different from 1, deviations from the Gaussian form occurs. 
When $\beta>0$ ($<0$) for 
$\nu>1$ ($<1$), associated to sub(super)-diffusion, 
clusters are platykurtic (leptokurtic). 
More importantly, according to Eq.~(\ref{eq:qgauss}), 
for $\beta>0$, clusters have the compact-support property 
(smooth boundary for $0<\beta<1$ and sharp for $\beta>1$).
This natural cutoff could in principle be associated to fragmentation. 
But, since clusters are not isolated, there is an additional condition for fragmentation: 
clusters should not overlap. 
This condition occurs when the support length is shorter than  the pattern wavelength, 
that is $2 x_0 < \Lambda$.
It is interesting to remark that these conditions match fairly well (not shown) the fragmentation region in the phase diagram of Fig.~\ref{fig:df}.

The agreement between the ansatz in Eq.~(\ref{eq:qgauss}) 
and numerical patterns opens an interesting question regarding the possibility of achieving 
an, at least approximate, analytical solution of Eq.~(\ref{maineq}), 
as found for linear processes~\cite{Barbosa2017}. 
Nevertheless, from direct substitution of the ansatz into Eq.~(\ref{maineq}), 
a straightforward result was not found. 
Moreover, the relation between the ansatz exponent $\beta$ 
and the  model exponents $\mu$, $\nu$ 
is not evident, but there is a strong trend given by $(\nu-1)$ 
(see Fig.~\ref{fig:beta}a). 
This major contribution to $\beta$  
corresponds to the exponent that emerges solely by nonlinear diffusion \cite{celia2005}.
Besides that, the exponent depends also on $\mu$ in a nontrivial way, as can be seen in 
Fig.~\ref{fig:beta}b.

{\bf Acknowledgments:} This work is partially funded by Brazilian Research Agencies Coordena\c c\~ao de Aperfei\c coamento de Pessoal de N\'ivel Superior (CAPES), Conselho Nacional de Desenvolvimento Cient\'ifico e Tecnol\'ogico (CNPq), Funda\c c\~ao de Amparo \`a Pesquisa do Estado do Rio de Janeiro (FAPERJ), and by the Spanish Research Agency, through grant MDM-2017-0711 from the Maria de Maeztu Program for units of Excellence in R\&D.

%

\bibliographystyle{apsrev4-1}

%

\end{document}